\documentclass[runningheads]{llncs}
\usepackage{amsfonts}
\usepackage{amsmath}
\usepackage{CJKutf8}
\usepackage[T1]{fontenc}
\usepackage{booktabs}
\usepackage{tabularx}
\usepackage{array}
\usepackage{graphicx}
\usepackage{xcolor}
\usepackage{listings}
\usepackage{enumitem}
\usepackage{amsmath}
\usepackage{amsfonts}
\usepackage{amssymb}
\usepackage{lmodern}
\usepackage{mathpazo}
\usepackage{url}
\usepackage{microtype} 
\usepackage{ragged2e}
\usepackage{hyperref}

\hypersetup{
    colorlinks=false,
    pdfborder={0 0 0},
    allbordercolors=0 0 0,
    linkbordercolor=0 0 0,
    citebordercolor=0 0 0,
    filebordercolor=0 0 0,
    urlbordercolor=0 0 0
}

\lstdefinestyle{promptblock}{
  basicstyle=\ttfamily\footnotesize,
  breaklines=true, breakatwhitespace=false,
  columns=fullflexible, keepspaces=true, showstringspaces=false,
  frame=single, framerule=0.4pt, rulecolor=\color{black!20},
  backgroundcolor=\color{black!2}
}

\usepackage{graphicx}
\usepackage{booktabs}
\usepackage{multirow}

\setlist[itemize]{leftmargin=*, topsep=3pt, partopsep=0pt, itemsep=2pt, parsep=0pt}
\setlist[enumerate]{leftmargin=*, topsep=3pt, partopsep=0pt, itemsep=2pt, parsep=0pt}

\newcolumntype{Y}{>{\centering\arraybackslash}X}
\newcolumntype{L}{>{\raggedright\arraybackslash}X}
\newcolumntype{R}{>{\raggedleft\arraybackslash}X}

\begin{document}

\title{CLLMRec: LLM-powered Cognitive-Aware Concept Recommendation via Semantic Alignment and Prerequisite Knowledge Distillation}
\titlerunning{CLLMRec: LLM-powered Cognitive-Aware Concept Recommendation}

\author{
Xiangrui Xiong\inst{1} \and
Yichuan Lu\inst{1} \and
Zifei Pan\inst{1} \and
Chang Sun\inst{1}
}

\authorrunning{X. Xiong et al.}

\institute{
Xihua University, Chengdu 610039, China\\
\email{whalealice1998@gmail.com}, 
\email{kotori\_itsuka@stu.xhu.edu.cn},
\email{asuka@stu.xhu.edu.cn},
\email{biluosection@gmail.com}\\
}

\maketitle

\begin{abstract}
The growth of Massive Open Online Courses (MOOCs) presents significant challenges for personalized learning, where concept recommendation is crucial. Existing approaches typically rely on heterogeneous information networks or knowledge graphs to capture conceptual relationships, combined with knowledge tracing models to assess learners' cognitive states. However, these methods face significant limitations due to their dependence on high-quality structured knowledge graphs, which are often scarce in real-world educational scenarios. To address this fundamental challenge, this paper proposes CLLMRec, a novel framework that leverages Large Language Models through two synergistic technical pillars: Semantic Alignment and Prerequisite Knowledge Distillation. The Semantic Alignment component constructs a unified representation space by encoding unstructured textual descriptions of learners and concepts. The Prerequisite Knowledge Distillation paradigm employs a teacher-student architecture, where a large teacher LLM (implemented as the Prior Knowledge Aware Component) extracts conceptual prerequisite relationships from its internalized world knowledge and distills them into soft labels to train an efficient student ranker. Building upon these foundations, our framework incorporates a fine-ranking mechanism that explicitly models learners' real-time cognitive states through deep knowledge tracing, ensuring recommendations are both structurally sound and cognitively appropriate. Extensive experiments on two real-world MOOC datasets demonstrate that CLLMRec significantly outperforms existing baseline methods across multiple evaluation metrics, validating its effectiveness in generating truly cognitive-aware and personalized concept recommendations without relying on explicit structural priors.
\keywords{Concept Recommendation \and Large Language Models \and Knowledge Distillation \and Recommender Systems \and Cognitive-Aware Recommendation}
\end{abstract}

\section{Introduction}
The deep integration of artificial intelligence and online education has transformed knowledge acquisition paradigms, creating an urgent need for intelligent personalized learning systems. Within this landscape, concept recommendation emerges as a cornerstone technology for constructing adaptive learning pathways, aiming to recommend the most suitable next knowledge concept for individual learners \cite{li2024learning,gligorea2023adaptive,gong2020attentional}.

Achieving high-quality concept recommendation necessitates addressing two fundamental challenges: accurately modeling learners' dynamic knowledge states and capturing the intricate prerequisite relationships among concepts. Traditional approaches typically employ knowledge tracing models to address the former \cite{zhang2024personalized,chen2023set}, while relying on external knowledge structures like knowledge graphs or heterogeneous information networks for the latter \cite{gong2020attentional,shi2020learning,zheng2024multigranularity}. However, these methods face significant limitations. First, over-emphasis on cognitive state indicators may overlook learners' genuine interests and deeper learning motivations \cite{li2024learning}. Second, the external knowledge structures they depend on often exhibit sparsity and static characteristics, struggling to adapt to dynamic learning environments where foundational concepts accumulate substantial interaction data while long-tail concepts suffer from sparsity \cite{yu2023graph}.

The emergence of large language models offers a transformative pathway to overcome these challenges. In general recommendation domains, LLMs have demonstrated remarkable capabilities in semantic alignment for enhancing item representations \cite{zhang2024notellm,lin2024bridging} and directly modeling complex user-item interactions \cite{zhu2024collaborative,lu2024aligning}. However, when applied to concept recommendation, their potential remains underexplored. Current research primarily focuses on using LLMs for generating static concept embeddings \cite{li2024learning}, failing to fully leverage their internalized world knowledge and reasoning capabilities to directly excavate and calibrate conceptual prerequisite relationships.

To address these limitations, we propose \textbf{CLLMRec}, a novel framework that systematically integrates Large Language Models through two core technical innovations: \textbf{Semantic Alignment} and \textbf{Prerequisite Knowledge Distillation}. The framework initiates with semantic alignment encoding, transforming unstructured textual descriptions of learners and concepts into a unified semantic space to establish robust foundational representations. Building upon this foundation, we introduce a novel prerequisite knowledge distillation paradigm where a large teacher LLM (implemented as the Prior Knowledge Aware Component, PKAC) activates its internalized world knowledge to perceive conceptual prerequisite relationships and generates distilled soft labels. These labels then guide the training of an efficient student ranker through a carefully designed distillation process. The student ranker further incorporates learners' historical interactions and preferences to generate personalized recommendations, while a fine-ranking module integrates real-time cognitive states for final calibration.

The main contributions of this work are threefold:
\begin{itemize}
    \item We propose CLLMRec, the first framework that integrates \textbf{Semantic Alignment} and \textbf{Prerequisite Knowledge Distillation} for concept recommendation, effectively modeling conceptual prerequisite relationships without requiring pre-constructed knowledge graphs.
    \item We design a novel \textbf{prerequisite knowledge distillation paradigm} that employs a teacher-student architecture, where knowledge about conceptual dependencies is extracted from large teacher LLMs and efficiently transferred to lightweight student models.
    \item We conduct extensive experiments on two real-world MOOC datasets, demonstrating that CLLMRec significantly outperforms existing baseline methods and provides valuable insights into LLM-powered educational recommendation.
\end{itemize}

\section{Related Work}

\subsection{Concept Recommendation}
Contemporary concept recommendation methodologies follow two primary trajectories: the simulator approach and the interest-based approach. The simulator approach centers around knowledge tracing models to construct simulated environments that replicate the dynamic evolution of learners' knowledge states, frequently integrated with reinforcement learning \cite{cai2019learning,li2023graph}. For instance, the SRC model \cite{chen2023set} incorporates deep knowledge tracing and collaborative knowledge tracing models, while the EK-TCP model \cite{he2022exercise} employs an enhanced deep knowledge tracing module to perceive prerequisite relationships. However, this approach's performance is heavily contingent on simulator accuracy and faces challenges in reward function design and training stability \cite{li2024learning}.

The interest-based approach directly utilizes behavioral sequences from authentic learning contexts as preference signals, reconstructing learning trajectories through sequence modeling and graph neural networks without intermediary simulation. Representative works include leveraging heterogeneous information networks or knowledge graphs to learn conceptual representations \cite{gong2020attentional,gong2023reinforced,wu2020exercise}. For example, ACKRec \cite{gong2020attentional} transforms implicit learner-concept associations into computable structural features through complex heterogeneous information networks. However, this approach's effectiveness is substantially constrained by the coverage and quality of external knowledge graphs.

\subsection{LLM-Enhanced Recommendation}
The application of large language models in recommendation systems encompasses three primary paradigms: (1) LLMs as semantic aligners that map unstructured textual information into unified semantic spaces to generate high-quality representations \cite{zhang2024notellm,wang2023generative}; (2) LLMs as external knowledge sources that generate auxiliary signals through prompt engineering \cite{wang2023enhancing,zeng2024federated}; and (3) LLMs as task models and distillation sources, either through direct fine-tuning on user-item interactions \cite{zhu2024collaborative} or by distilling LLM capabilities into lightweight recommendation networks \cite{cui2024distillation}.

While LLMs have been introduced to concept recommendation, existing research predominantly focuses on generating static concept representations \cite{li2024learning}, failing to systematically leverage LLMs' world knowledge and reasoning capabilities for direct prerequisite relationship mining. Our work bridges this gap by proposing a comprehensive framework that integrates semantic alignment with a novel prerequisite knowledge distillation paradigm.

\section{The CLLMRec Framework}

\subsection{Preliminaries}
We formalize the concept recommendation task as follows: given a set of learners $U=\{u_1, u_2, \dots, u_N\}$ and a set of concepts $K=\{k_1, k_2, \dots, k_M\}$, at any time step $t$, the model recommends a list of most suitable subsequent learning concepts for learner $u$ based on their historical interaction sequence $H_u^t = (k_1, k_2, \dots, k_{t-1})$.

\subsection{Framework Overview}
\begin{figure}[htbp]
    \centering
    \includegraphics[trim={0.2cm 0 0.1cm 0}, clip, width=\linewidth]{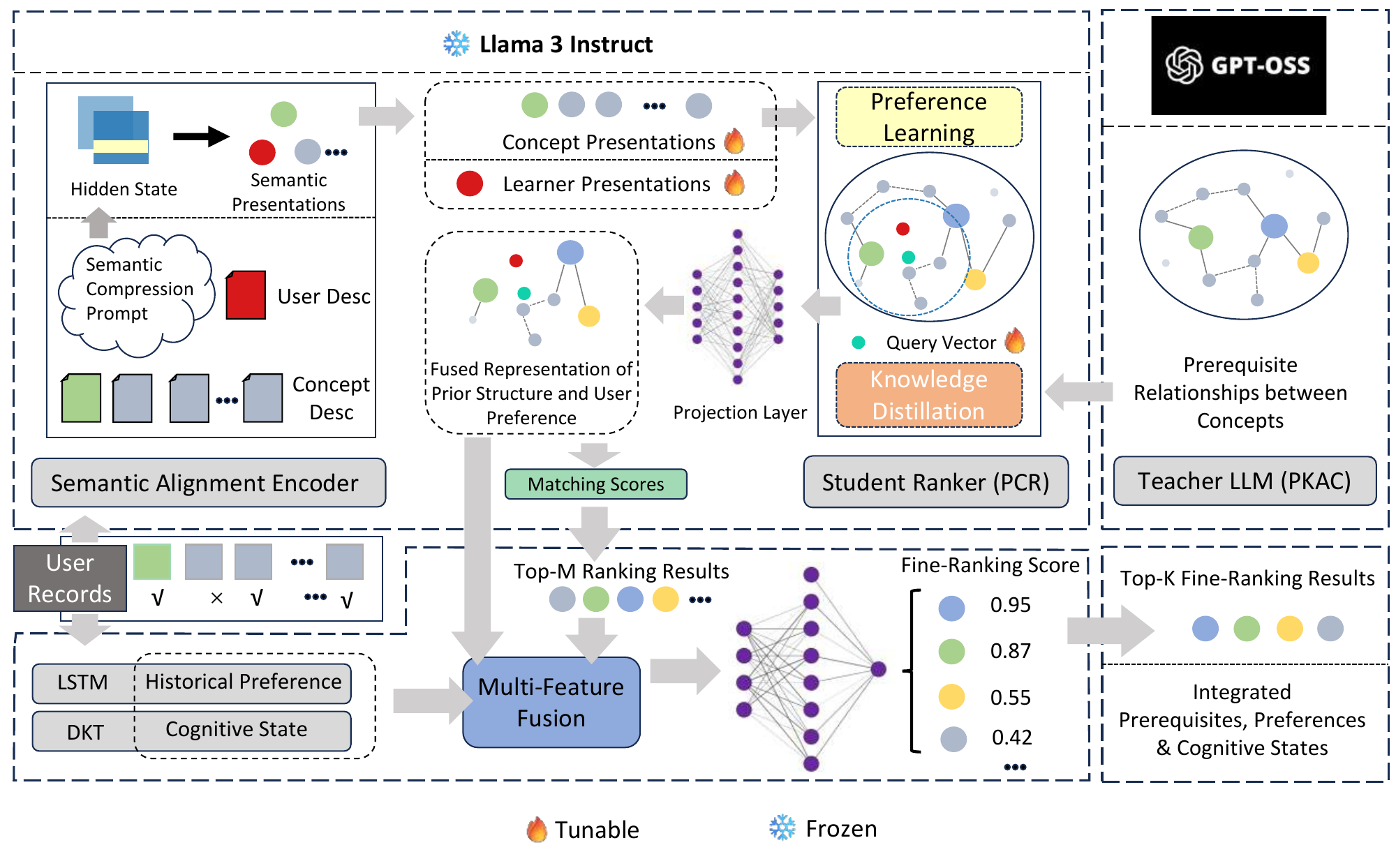}
    \caption{Overall Architecture and Data Flow of CLLMRec. The diagram illustrates the end-to-end pipeline: user and concept data are encoded by the LLM, processed by the ranker which fuses prerequisites and preferences, and finally calibrated with cognitive states to produce recommendations.}
    \label{fig:CLLMRec-architecture}
\end{figure}

CLLMRec is architected around two core technical innovations: \textbf{Semantic Alignment} and \textbf{Prerequisite Knowledge Distillation}. As illustrated in Figure~\ref{fig:CLLMRec-architecture}, the framework comprises four key components that work synergistically: (1) a Semantic Alignment Encoder that constructs unified representations from unstructured texts; (2) a Teacher LLM for Prerequisite Knowledge Distillation (implemented as the Prior Knowledge Aware Component) that extracts conceptual dependency relationships; (3) a Student Ranker that learns through knowledge distillation and preference modeling; and (4) a Fine-Ranker that incorporates cognitive states for final calibration.

The framework operates through a coherent multi-stage workflow: semantic alignment encoding transforms textual descriptions into unified embeddings; prerequisite knowledge distillation employs a teacher-student paradigm to transfer knowledge about conceptual dependencies; personalized coarse ranking generates initial recommendations; and fine-ranking calibration incorporates cognitive states for final optimization.

\subsection{Large Language Model Encoder}
In CLLMRec, we employ frozen LLM parameters, utilizing only the forward encoding capability. We define the model in this mode as encoder \(f_\theta(\cdot)\). Given an input prompt \(\mathrm{p}\), we extract anchor token embeddings from the last hidden layer through function \(f_a(T, f_\theta(\mathrm{p}))\), where \(T\) denotes the specific anchor token (e.g., `[C]` for concepts or `[S]` for students) used to extract the semantic representation.

\subsection{Semantic Alignment Encoder: Unified Semantic Representations}
The Semantic Alignment Encoder addresses the challenge of heterogeneous textual information in educational platforms by transforming unstructured text into unified, retrievable vector representations. This component employs carefully designed prompt templates (Table~\ref{tab:assistant-prompts}) with semantic anchors (`[C]` for concepts, `[S]` for students) to guide the LLM in extracting core semantics from verbose descriptions. Anchor tokens are special tokens strategically placed in prompts to serve as semantic carriers, enabling precise extraction of relevant information from the LLM's hidden states.

\begin{table}[htbp]
  \centering
  \caption{Prompt templates for Semantic Alignment}
  \label{tab:assistant-prompts}
  \renewcommand{\arraystretch}{1.2}
  \begin{tabularx}{0.9\linewidth}{@{}p{2.5cm} X@{}}
    \toprule
    \textbf{Entity Type} & \textbf{Prompt with Semantic Anchor} \\
    \midrule
    Concept &
    \begin{minipage}[t]{\linewidth}\ttfamily\small
Extract the concept's description information and compress it into one word for recommendation.\\
The description is: <description>.\\
The compression word is: '[C]'.
    \end{minipage}
    \\[0.8em]
    \midrule
    Student &
    \begin{minipage}[t]{\linewidth}\ttfamily\small
Extract the student's profile information and compress it into one word for identification.\\
The description is: <description>.\\
The compression word is: '[S]'.
    \end{minipage}
    \\
    \bottomrule
  \end{tabularx}
\end{table}

Through \(f_a(T, f_\theta(\mathrm{p_a}))\) and \(\ell_2\) normalization, we obtain semantic embeddings \(\mathbf{e}\) for each entity, resulting in user embedding matrix \(\mathbf{E} \in \mathbb{R}^{N \times d}\) and concept embedding matrix \(\mathbf{C} \in \mathbb{R}^{M \times d}\). This semantic alignment provides the foundational representation space for subsequent components.

\subsection{Prerequisite Knowledge Distillation: Teacher-Student Paradigm}
The Prerequisite Knowledge Distillation component addresses the critical limitation of dependency on external knowledge graphs by employing a teacher-student architecture. The teacher LLM, implemented as the \textbf{Prior Knowledge Aware Component}, activates its internalized world knowledge to perceive conceptual dependency relationships, while the student model (Personalized Coarse Ranker) learns to replicate this knowledge through distillation.

\subsubsection{Prior Knowledge Aware Component (Teacher LLM): Soft Label Generation}
The Prior Knowledge Aware Component employs a structured prompt system (Table~\ref{tab:expert-prompts}) to generate integer scores \( a_j \in [s_{\min}, s_{\max}] \cap \mathbb{Z} \) for each candidate concept, which are then transformed into probability distributions through normalization and smoothing:

\begin{equation}
p_j = \frac{\max(0, a_j/\max(1, \max_k a_k))}{\sum_k \max(0, a_k/\max(1, \max_k a_k))}, \quad 
y_j^{(e)} = (1 - \varepsilon) p_j + \frac{\varepsilon}{M}
\end{equation}
where \(\varepsilon\) is the label smoothing factor to prevent overconfidence. These soft labels \(\mathbf{y}^{(e)}\) encapsulate the teacher LLM's understanding of conceptual prerequisite relationships and serve as the supervisory signal for distillation.

\begin{table}[htb]
  \centering
  \caption{Prompt structure for Prerequisite Knowledge Distillation}
  \label{tab:expert-prompts}
  \renewcommand{\arraystretch}{1.2}
  \begin{tabularx}{0.9\linewidth}{@{}p{3cm} X@{}}
    \toprule
    \textbf{Component} & \textbf{Content} \\
    \midrule
    Task Prompt  &
    \begin{minipage}[t]{\linewidth}\ttfamily\small
You are an expert learning planner.\\
Return ONLY a valid JSON object exactly in the following schema:\\
\{ "scores": [ \{ "id": <int>, "score": <int> \} , ... ] \}

Hard rules:\\
- Score ONLY the concepts listed in the provided chunk.\\
- The target concept is the final goal; DO NOT recommend the target itself.\\
- Use integer scores in the closed interval [<score\_min>, <score\_max>].\\
- JSON only. No extra text, no markdown, no explanations.
    \end{minipage}
    \\[0.8em]
    \midrule
    Task Data  &
    \begin{minipage}[t]{\linewidth}\ttfamily\small
\{\\
\quad "target": "<target concept text>",\\
\quad "history": [ "<history text1>", "<history text2>", "..." ],\\
\quad "concept\_chunk": [ \{ "id": 101, "concept": "<candidate concept text>" \}, ... ],\\
\quad "score\_scale": \{ "min": 0, "max": 3 \}\\
\}
    \end{minipage}
    \\
    \bottomrule
  \end{tabularx}
\end{table}

\subsubsection{Student Ranker: Knowledge Acquisition}
The student model (Personalized Coarse Ranker, PCR) comprises a frozen LLM encoder and a learnable projection layer. It generates query vectors using prompt templates (Table~\ref{tab:teacher-prompts}) and computes matching scores through:
\begin{equation}
\mathbf{s}_j = \phi(\mathbf{q}_w, \mathbf{e}_u, \mathbf{c}) = \mathbf{q}_w \cdot \mathbf{c}_j^T + \alpha \cdot \mathbf{e}_u \cdot \mathbf{c}_j^T,
\end{equation}
where \( \mathbf{q}_w \) is the query vector, \( \mathbf{c}_j \) is the concept embedding, and \( \mathbf{e}_u \) is the learner embedding. Here, \(\alpha\) is a trainable weight parameter that balances the contribution of user preference against query-concept relevance.

\begin{table}[htbp]
  \centering
  \caption{Knowledge distillation prompts for student model}
  \label{tab:teacher-prompts}
  \renewcommand{\arraystretch}{1.2}
  \begin{tabularx}{0.9\linewidth}{@{}p{5cm} X@{}}
    \toprule
    \textbf{Stage} & \textbf{Prompt} \\
    \midrule
    Knowledge Distillation &
    \begin{minipage}[t]{\linewidth}\ttfamily\small
History: <hist\_1> | <hist\_2> | ... \\
Target: <target\_text> \\
Recommend the next concept:
    \end{minipage}
    \\[0.8em]
    \midrule
    Preference Learning &
    \begin{minipage}[t]{\linewidth}\ttfamily\small
History: <hist\_1> | <hist\_2> | ... \\
Recommend the next concept:
    \end{minipage}
    \\
    \bottomrule
  \end{tabularx}
\end{table}

The distillation process occurs in two stages:

\textbf{Knowledge Distillation Stage}: The student model learns to mimic the teacher's decision distribution by minimizing:
\begin{equation}
\mathcal{L}_{\text{distill}} = -\sum_{j=1}^{M} y_j^{(e)} \log\left(\sigma\left(\frac{\mathbf{s}}{\tau}\right)\right)_j,
\end{equation}
where \( y_j^{(e)} \) are the teacher-generated soft labels and \( \tau \) is a temperature parameter.

\textbf{Preference Learning Stage}: The student model incorporates personalized preferences through contrastive learning:
\begin{equation}
\mathcal{L}_{pref} = -\log \frac{\exp(\phi(\mathbf{q}_w, \mathbf{e}_u, \mathbf{c}^+))}{\exp(\phi(\mathbf{q}_w, \mathbf{e}_u, \mathbf{c}^+)) + \sum_{j=1}^{K} \exp(\phi(\mathbf{q}_w, \mathbf{e}_u, \mathbf{c}_j^-))}.
\end{equation}

\subsection{Fine-Ranker: Cognitive State Integration}
The Fine-Ranker incorporates real-time cognitive states for final recommendation calibration. It computes the final score $s'(c)$ by fusing four matching features through an MLP:
\begin{equation}
s'(c) = \text{MLP}\left(f_{\text{user}}(c) \oplus f_{\text{hist}}(c) \oplus f_{\text{coarse}}(c) \oplus f_{\text{dkt}}(c)\right)
\end{equation}
where the features are computed as follows:
\begin{itemize}
    \item $f_{\text{user}}(c)$: user-concept affinity via dot product between projected user and concept embeddings
    \item $f_{\text{hist}}(c)$: historical relevance via dot product between LSTM-encoded historical sequence and concept embeddings  
    \item $f_{\text{coarse}}(c)$: coarse ranking score from the student ranker
    \item $f_{\text{dkt}}(c)$: The feature $f_{\text{dkt}}(c)$ is designed to calibrate the cognitive difficulty of recommendations. It computes a matching score via the dot product between the cognitive state vector from the DKT\cite{piech2015deep} module (after a projection) and the candidate concept's embedding, thereby measuring the alignment between the user's current knowledge state and the conceptual prerequisite requirements.
\end{itemize}
The model is optimized using pairwise ranking loss with candidate coverage handling:
\begin{equation}
\mathcal{L}_{\text{reranker}} = -\frac{1}{|\mathcal{B}|} \sum_{u \in \mathcal{B}} \mathbb{I}(c^+_u \in \mathcal{C}_u) \cdot \log \frac{\exp(s'(c^+_u))}{\exp(s'(c^+_u)) + \sum_{c \in \mathcal{N}_u} \exp(s'(c))}
\end{equation}

This design ensures that the fine-ranker effectively leverages both semantic information from the embeddings and cognitive state information from the DKT module, while maintaining computational efficiency through candidate pre-selection and feature projection.

\subsection{Framework Integration}
The four components of CLLMRec work synergistically: Semantic Alignment provides unified representations, Prerequisite Knowledge Distillation transfers structural knowledge, the Student Ranker generates personalized recommendations, and the Fine-Ranker incorporates cognitive states for final optimization. 

The integrated optimization objective of CLLMRec combines the learning signals from all components:
\begin{equation}
\mathcal{L}_{\text{total}} = \lambda_1 \mathcal{L}_{\text{distill}} + \lambda_2 \mathcal{L}_{\text{pref}} + \lambda_3 \mathcal{L}_{\text{reranker}}
\end{equation}
where $\lambda_1$, $\lambda_2$, and $\lambda_3$ are balancing coefficients that control the relative importance of knowledge distillation, preference learning, and fine-ranking objectives respectively.

This integrated approach enables CLLMRec to generate conceptually sound and highly personalized recommendations without dependency on external knowledge graphs, while maintaining a coherent end-to-end learning framework.

\section{Experiments}

\subsection{Experimental Setup}

\subsubsection{Datasets}
We evaluate CLLMRec on two public MOOC datasets: ASSIST09 \cite{feng2009addressing} and ASSIST12 \cite{pardos2013affective}. Both datasets lack explicit prerequisite structures, providing an ideal testbed for evaluating our framework's ability to infer conceptual relationships without external knowledge. Dataset statistics are shown in Table~\ref{tab:dataset}.

\begin{table}[htbp]
    \centering
    \caption{Dataset statistics}
    \label{tab:dataset}
    \begin{tabular}{lcccc}
        \toprule
        \textbf{Dataset} & \textbf{Learners} & \textbf{Concepts} & \textbf{Interactions} & \textbf{Avg. Sequence Length} \\
        \midrule
        ASSIST09 & 3,322 & 145 & 187,914 & 56.6 \\
        ASSIST12 & 24,155 & 265 & 1,853,338 & 76.7 \\
        \bottomrule
    \end{tabular}
\end{table}

\subsubsection{Implementation Details}
We employ Llama3-8B-Instruct as the base encoder for semantic alignment and student ranking, and GPT-OSS-20B as the teacher model for prerequisite knowledge distillation. All experiments are conducted using PyTorch with Adam optimizer, and we report the average results over five random seeds.

\subsubsection{Baseline Methods}
To comprehensively evaluate CLLMRec's performance, we compare it against five representative baseline methods spanning different recommendation paradigms:

\begin{itemize}
    \item \textbf{ACKRec} \cite{gong2020attentional}: Models learner-concept relationships via meta-path guided graph convolution on heterogeneous information networks.
    
    \item \textbf{GCARec} \cite{yu2023graph}: A framework leveraging graph contrastive learning and generative adversarial networks to address data sparsity by generating virtual representations for long-tail concepts.
    
    \item \textbf{BERT4Rec} \cite{sun2019bert4rec}: A sequence recommendation model based on bidirectional Transformer, capturing long-term dependencies in users' interaction sequences with self-attention.
    
    \item \textbf{Random}: A baseline recommendation strategy using uniform random selection.
    
    \item \textbf{Rule-based}: A rule-based approach combined with a deep knowledge tracing (DKT) model, recommending the least familiar concepts based on cosine similarity to users' cognitive states.
\end{itemize}

\subsubsection{Evaluation Metrics}
We employ three standard evaluation metrics: Hit Rate (HR@K), Normalized Discounted Cumulative Gain (NDCG@K), and Mean Reciprocal Rank (MRR@K). We report HR@1 for measuring accuracy at the first position, and NDCG@5 and MRR@5 for assessing overall ranking quality within the top-5 recommendations.

\subsection{Results and Analysis}

\subsubsection{Performance Comparison}
As shown in Table~\ref{tab:performance-comparison}, CLLMRec significantly outperforms all baseline methods across both datasets and all evaluation metrics. On ASSIST09, CLLMRec achieves a HR@1 of 0.6359, representing a 153\% relative improvement over the best baseline (GCARec at 0.2513). This substantial performance gain demonstrates the effectiveness of our semantic alignment and knowledge distillation approach.

\begin{table}[htbp]
  \centering
  \caption{Performance comparison across datasets}
  \label{tab:performance-comparison}
  \begin{tabular}{lccccccc}
    \toprule
    \textbf{Dataset} & \textbf{Metric} & \textbf{ACKRec} & \textbf{GCARec} & \textbf{BERT4Rec} & \textbf{Random} & \textbf{Rule-based} & \textbf{CLLMRec} \\
    \midrule
    \multirow{3}{*}{ASSIST09} & HR@1 & 0.2408 & 0.2513 & 0.1935 & 0.0021 & 0.0027 & \textbf{0.6359} \\
                              & NDCG@5 & 0.3787 & 0.3904 & 0.2203 & 0.0061 & 0.0084 & \textbf{0.7031} \\
                              & MRR@5 & 0.3727 & 0.3756 & 0.2342 & 0.0047 & 0.0065 & \textbf{0.7290} \\
    \midrule
    \multirow{3}{*}{ASSIST12} & HR@1 & 0.1976 & 0.2119 & 0.2540 & 0.0009 & 0.0005 & \textbf{0.5365} \\
                              & NDCG@5 & 0.3078 & 0.3169 & 0.4247 & 0.0027 & 0.0012 & \textbf{0.6009} \\
                              & MRR@5 & 0.3064 & 0.3182 & 0.4047 & 0.0021 & 0.0017 & \textbf{0.5825} \\
    \bottomrule
  \end{tabular}
\end{table}

The performance advantages are consistent across datasets, with CLLMRec achieving 0.5365 HR@1 on ASSIST12, significantly outperforming BERT4Rec (0.2540). These results validate that semantic alignment and prerequisite knowledge distillation effectively address the limitations of traditional methods that depend on external knowledge structures.

\subsubsection{Ablation Study}
We conduct comprehensive ablation studies to evaluate the contribution of each component (Table~\ref{tab:ablation-study}). The CLLMRec-Only-Ranker variant demonstrates exceptional performance in knowledge distillation evaluation (0.9900 HR@1 on ASSIST09) but poor performance in preference learning evaluation (0.0320 HR@1), confirming that the knowledge distillation stage successfully transfers prerequisite knowledge while the preference learning stage incorporates personalized preferences.

\begin{table}[htbp]
    \centering
    \caption{Ablation study results}
    \label{tab:ablation-study}
    \begin{tabular}{lcccccc}
        \toprule
        \textbf{Dataset} & \textbf{Metric} & \textbf{Only-Student} & \textbf{w/o-DKT} & \multicolumn{2}{c}{\textbf{Only-Ranker}} & \textbf{Full Model} \\
        \cmidrule(lr){5-6}
        & & & & \textbf{K.D.} & \textbf{P.L.} & \\
        \midrule
        \multirow{3}{*}{ASSIST09} 
            & HR@1    & 0.1760 & 0.5095 & 0.9900 & 0.0320 & \textbf{0.6359} \\
            & NDCG@5  & 0.3981 & 0.6165 & 0.9860 & 0.0912 & \textbf{0.7031} \\
            & MRR@5   & 0.2956 & 0.5866 & 0.9950 & 0.0548 & \textbf{0.7290} \\
        \midrule
        \multirow{3}{*}{ASSIST12} 
            & HR@1    & 0.1683 & 0.4934 & 0.9200 & 0.0040 & \textbf{0.5365} \\
            & NDCG@5  & 0.3625 & 0.5426 & 0.9482 & 0.0400 & \textbf{0.6009} \\
            & MRR@5   & 0.2631 & 0.5282 & 0.9567 & 0.0163 & \textbf{0.5825} \\
        \bottomrule
    \end{tabular}
\end{table}

The CLLMRec-w/o-DKT variant shows significant performance degradation, highlighting the importance of cognitive state integration in the fine-ranking stage. This confirms that both prerequisite knowledge distillation and cognitive state awareness are essential for optimal concept recommendation.

\section{Conclusion}
This paper addresses the fundamental challenge of generating cognitively-aware concept recommendations without relying on explicit knowledge structures. We propose CLLMRec, a novel framework that integrates Large Language Models through two core technical innovations: Semantic Alignment and Prerequisite Knowledge Distillation. The Semantic Alignment component constructs unified representations from unstructured textual descriptions, while the Prerequisite Knowledge Distillation paradigm employs a teacher-student architecture to transfer knowledge about conceptual dependencies from large teacher LLMs to efficient student models.

Extensive experiments demonstrate that CLLMRec significantly outperforms existing baseline methods, achieving 153\% relative improvement in HR@1 on the ASSIST09 dataset. Ablation studies confirm the critical importance of both semantic alignment and knowledge distillation components. Our work establishes a new paradigm for educational recommendation that leverages LLMs' internalized knowledge rather than depending on external structural priors.

\bibliographystyle{splncs04}
\bibliography{references}

@article{li2024learning,
  title={Learning structure and knowledge aware representation with large language models for concept recommendation},
  author={Li, Q. and Xia, W. and Du, K. and Zhang, Q. and Zhang, W. and Tang, R. and Yu, Y.},
  journal={arXiv preprint arXiv:2405.12442},
  year={2024}
}

@article{gligorea2023adaptive,
  title={Adaptive learning using artificial intelligence in e-learning: a literature review},
  author={Gligorea, I. and Cioca, M. and Oancea, R. and Gorski, A.-T. and Gorski, H. and Tudorache, P.},
  journal={Educ. Sci.},
  volume={13},
  number={12},
  pages={1216--1238},
  year={2023},
  doi={10.3390/educsci13121216}
}

@article{zhang2024personalized,
  title={Personalized process-type learning path recommendation based on process mining and deep knowledge tracing},
  author={Zhang, F. and Feng, X. and Wang, Y.},
  journal={Knowl.-Based Syst.},
  volume={303},
  pages={112431},
  year={2024},
  doi={10.1016/j.knosys.2024.112431}
}

@inproceedings{gong2020attentional,
  title={Attentional graph convolutional networks for knowledge concept recommendation in MOOCs in a heterogeneous view},
  author={Gong, J. and Wang, S. and Wang, J. and Feng, W. and Peng, H. and Tang, J. and Yu, P.S.},
  booktitle={Proc. 43rd Int. ACM SIGIR Conf. Res. Dev. Inf. Retr.},
  pages={79--88},
  year={2020},
  doi={10.1145/3397271.3401157}
}

@article{shi2020learning,
  title={A learning path recommendation model based on a multidimensional knowledge graph framework for e-learning},
  author={Shi, D. and Wang, T. and Xing, H. and Xu, H.},
  journal={Knowl.-Based Syst.},
  volume={195},
  pages={105618},
  year={2020},
  doi={10.1016/j.knosys.2020.105618}
}

@article{zheng2024multigranularity,
  title={A multigranularity learning path recommendation framework based on knowledge graph and improved ant colony optimization algorithm for e-learning},
  author={Zheng, Y. and Wang, D. and Xu, Y. and Li, Y.},
  journal={IEEE Trans. Comput. Soc. Syst.},
  volume={11},
  number={2},
  pages={892--903},
  year={2024},
  doi={10.1109/TCSS.2023.3323649}
}

@inproceedings{zhang2024notellm,
  title={Notellm: a retrievable large language model for note recommendation},
  author={Zhang, C. and Wu, S. and Zhang, H. and Xu, T. and Gao, Y. and Hu, Y. and Chen, E.},
  booktitle={Companion Proc. ACM Web Conf.},
  pages={170--179},
  year={2024},
  doi={10.1145/3589334.3645109}
}

@inproceedings{lin2024bridging,
  title={Bridging items and language: a transition paradigm for large language model-based recommendation},
  author={Lin, X. and Wang, W. and Li, Y. and Feng, F. and Ng, S.-K. and Chua, T.-S.},
  booktitle={Proc. 30th ACM SIGKDD Conf. Knowl. Discov. Data Min.},
  pages={1816--1826},
  year={2024},
  doi={10.1145/3637528.3671888}
}

@inproceedings{zhu2024collaborative,
  title={Collaborative large language model for recommender systems},
  author={Zhu, Y. and Wu, L. and Guo, Q. and Hong, L. and Li, J.},
  booktitle={Proc. ACM Web Conf.},
  pages={3162--3172},
  year={2024},
  doi={10.1145/3589334.3645536}
}

@inproceedings{lu2024aligning,
  title={Aligning large language models for controllable recommendations},
  author={Lu, W. and Lian, J. and Zhang, W. and Li, G. and Zhou, M. and Liao, H. and Xie, X.},
  booktitle={Proc. 62nd Annu. Meet. Assoc. Comput. Linguist.},
  pages={8159--8172},
  year={2024},
  doi={10.18653/v1/2024.acl-long.454}
}

@inproceedings{cai2019learning,
  title={Learning path recommendation based on knowledge tracing model and reinforcement learning},
  author={Cai, D. and Zhang, Y. and Dai, B.},
  booktitle={Proc. 5th IEEE Int. Conf. Comput. Commun.},
  pages={1881--1885},
  year={2019},
  doi={10.1109/ICCC47050.2019.8944107}
}

@inproceedings{li2023graph,
  title={Graph enhanced hierarchical reinforcement learning for goal-oriented learning path recommendation},
  author={Li, Q. and Xia, W. and Yin, L. and Shen, J. and Rui, R. and Zhang, W. and Chen, X. and Tang, R. and Yu, Y.},
  booktitle={Proc. 32nd ACM Int. Conf. Inf. Knowl. Manag.},
  pages={1318--1327},
  year={2023},
  doi={10.1145/3583780.3614843}
}

@article{gong2023reinforced,
  title={Reinforced MOOCs concept recommendation in heterogeneous information networks},
  author={Gong, J. and Wan, Y. and Liu, Y. and Li, X. and Zhao, Y. and Wang, C. and Lin, Y. and Fang, X. and Feng, W. and Zhang, J. and others},
  journal={ACM Trans. Web},
  volume={17},
  number={3},
  pages={1--27},
  year={2023},
  doi={10.1145/3584818}
}

@article{wu2020exercise,
  title={Exercise recommendation based on knowledge concept prediction},
  author={Wu, Z. and Li, M. and Tang, Y. and Liang, Q.},
  journal={Knowl.-Based Syst.},
  volume={210},
  pages={106481},
  year={2020},
  doi={10.1016/j.knosys.2020.106481}
}

@article{wang2023generative,
  title={Generative recommendation: towards next-generation recommender paradigm},
  author={Wang, W. and Lin, X. and Feng, F. and He, X. and Chua, T.-S.},
  journal={arXiv preprint arXiv:2304.03516},
  year={2023}
}

@article{wang2023enhancing,
  title={Enhancing recommender systems with large language model reasoning graphs},
  author={Wang, Y. and Chu, Z. and Ouyang, X. and Wang, S. and Hao, H. and Shen, Y. and Gu, J. and Xue, S. and Zhang, J.Y. and Cui, Q. and others},
  journal={arXiv preprint arXiv:2308.10835},
  year={2023}
}

@inproceedings{zeng2024federated,
  title={Federated recommendation via hybrid retrieval augmented generation},
  author={Zeng, H. and Yue, Z. and Jiang, Q. and Wang, D.},
  booktitle={Proc. IEEE Int. Conf. Big Data},
  pages={8078--8087},
  year={2024},
  doi={10.1109/BigData59044.2024.10682145}
}

@inproceedings{piech2015deep,
  title={Deep knowledge tracing},
  author={Piech, C. and Bassen, J. and Huang, J. and Ganguli, S. and Sahami, M. and Guibas, L.J. and Sohl-Dickstein, J.},
  booktitle={Adv. Neural Inf. Process. Syst.},
  pages={505--513},
  year={2015}
}

@inproceedings{yu2023graph,
  title={Graph contrastive learning with adaptive augmentation for knowledge concept recommendation},
  author={Yu, M. and Ding, Z. and Yu, J. and Zhang, W. and Yang, M. and Zhao, M.},
  booktitle={Proc. 26th Int. Conf. Comput. Supported Coop. Work Des.},
  pages={1281--1286},
  year={2023}
}

@article{he2022exercise,
  title={Exercise recommendation method based on knowledge tracing and concept prerequisite relations},
  author={He, Y. and Wang, H. and Pan, Y. and Zhou, Y. and Sun, G.},
  journal={CCF Trans. Pervasive Comput. Interact.},
  volume={4},
  number={4},
  pages={452--464},
  year={2022}
}

@inproceedings{chen2023set,
  title={Set-to-sequence ranking-based concept-aware learning path recommendation},
  author={Chen, X. and Shen, J. and Xia, W. and Jin, J. and Song, Y. and Zhang, W. and Liu, W. and Zhu, M. and Tang, R. and Dong, K. and others},
  booktitle={Proc. AAAI Conf. Artif. Intell.},
  volume={37},
  number={4},
  pages={5027--5035},
  year={2023}
}

@inproceedings{cui2024distillation,
  title={Distillation matters: empowering sequential recommenders to match the performance of large language models},
  author={Cui, Y. and Liu, F. and Wang, P. and Wang, B. and Tang, H. and Wan, Y. and Wang, J. and Chen, J.},
  booktitle={Proc. 18th ACM Conf. Recommender Syst.},
  pages={507--517},
  year={2024}
}

@inproceedings{pardos2013affective,
  title={Affective states and state tests: investigating how affect throughout the school year predicts end of year learning outcomes},
  author={Pardos, Z.A. and Baker, R.S.J.D. and San Pedro, M.O.C.Z. and Gowda, S.M. and Gowda, S.M.},
  booktitle={Proc. 3rd Int. Conf. Learn. Anal. Knowl.},
  pages={117--124},
  year={2013}
}

@article{feng2009addressing,
  title={Addressing the assessment challenge with an online system that tutors as it assesses},
  author={Feng, M. and Heffernan, N. and Koedinger, K.},
  journal={User Model. User-Adapt. Interact.},
  volume={19},
  number={3},
  pages={243--266},
  year={2009}
}

@inproceedings{sun2019bert4rec,
  title={BERT4Rec: sequential recommendation with bidirectional encoder representations from transformer},
  author={Sun, F. and Liu, J. and Wu, J. and Pei, C. and Lin, X. and Ou, W. and Jiang, P.},
  booktitle={Proc. 28th ACM Int. Conf. Inf. Knowl. Manag.},
  pages={1441--1450},
  year={2019}
}
\end{document}